%% file: OCCitFoASaL.tex
\documentclass{article}

\usepackage{url}
\usepackage{listings}

\usepackage{common}

\newtheorem{assumption}{Assumption}[section]

\newcommand{\causes}{\,{\rightarrow}\,}

\newcommand{\squiggly}{\!\!\xymatrix@C=1em{{}\ar@{~>}[r]&{}}\!\!}
\newcommand{\lcorner}{\raisebox{-2.5pt}{$\llcorner{}\hspace{-2pt}$}}
\newcommand{\xlcorner}{\raisebox{-4.5pt}{$\llcorner{}\hspace{-2pt}$}}
\newcommand{\rcorner}{\raisebox{-2.5pt}{$\hspace{-2pt}\lrcorner{}$}}
\newcommand{\xrcorner}{\raisebox{-2.5pt}{$\hspace{-4pt}\lrcorner{}$}}
\newcommand{\lsep}{<\!\!<}
\newcommand{\rsep}{>\!\!>}
\newcommand{\llbrack}{\lbrack\hspace{-0.15em}\lbrack}
\newcommand{\rrbrack}{\rbrack\hspace{-0.15em}\rbrack}

\begin{document}

\title{\normalfont\bfseries\large Operational Concurrency Control\\in the Face of Arbitrary Scale and Latency}
\author{James Smith\\\texttt{james.smith@djalbat.com}}
\date{}
	
\maketitle

\begin{abstract}
\noindent We present for the first time a complete solution to the problem of proving the correctness of a concurrency control algorithm for collaborative text editors against the standard consistency model. The success of our approach stems from the use of com- prehensive stringwise operational transformations, which appear to have escaped a formal treatment until now. Because these transformations sometimes lead to an increase in the number of operations as they are transformed, we cannot use inductive methods and adopt the novel idea of decreasing diagrams instead. We also base our algorithm on a client-server model rather than a peer-to-peer one, which leads to the correct application of operational transformations to both newly generated and pending operations. And lastly we solve the problem of latency, so that our algorithm works perfectly in practice. The result of these innovations is the first ever formally correct concurrency control algorithm for collaborative text editors together with a fast, fault tolerant and highly scalable implementation.
\end{abstract}

\include*{introduction}
\include*{operationaltransformations}
\include*{protocol}

\include*{consistency}
\include*{conclusions}

\pagebreak

\bibliographystyle{plain}
\bibliography{references}

\end{document}

%% file: introduction.tex
\section{Introduction}

Collaborative text editors have something of a convoluted history. The idea was first publicly mooted in 1968 by Turing Award winner Douglas Engelbart in his landmark demo that posthumously became known as ``The Mother of all Demos''~\cite{Engelbart}. Some twenty years later the first paper on a concurrency control algorithm appeared, although no correctness proof was given~\cite{Ellis:1989:CCG:66926.66963}. Indeed the algorithm was found to be incorrect and partial alternatives were proposed, this time along with correctness proofs~\cite{Cormack:1995:CCU:224964.225007,Ressel:1996:ITA:240080.240305}. The standard consistency model was also defined around this time~\cite{Sun96aconsistency}. However, in spite of the plethora of algorithms and implementations that followed, the problem of proving the correctness of a concurrency control algorithm against this consistency model seems never to have been solved.

We briefly outline some of the issues behind this. To begin with, formally correct characterwise operational transformations remained elusive until relatively recently~\cite{imine2003proving} whilst formally correct stringwise operational transformations cannot be found in the literature at all. We think the unreasonable correctness criteria put upon operational transformations by peer-to-peer algorithms in particular are part of the reason for this. In fact despite their inherent complexity peer-to-peer algorithms have nearly always been preferred~\cite{Vidot:2000:CCD:358916.358988,oster:inria-00071240,10.1109/TPDS.2009.173} and have persisted up until the present. Other consistency models have now also been proposed~\cite{li2005commutativity,DBLP:journals/tpds/LiL07,sun2009context,
letia2009crdts,journals/cscw/LiL10}, no doubt in response to an inability to prove the correctness of any algorithm against the standard one, and these cloud the picture. Lastly, modern collaborative platforms such as Google Docs require complex data types other than plain text, and this complicates things still further.

Our solution addresses all of these issues. Firstly, we give a common-sense and comprehensive definition of stringwise operational transformations, we think for the first time. Secondly, we base our algorithm on the client-server model, which we feel is more appropriate to a modern Internet setting. Thirdly, when we prove correctness against a consistency model, we do so against the standard one, which, if a consistency model is needed at all, is adequate. Lastly, we work only with plain text documents and their attendant inserts and deletes, but solve this problem completely.

In what follows, for the most part we give the details of our algorithm and demonstrate its correctness \emph{first}, before outlining the concepts and contributions to be found elsewhere. Our reason is this: our algorithm was conceived in a vacuum, so to speak, without knowledge of the surrounding literature, and we think that this approach contributed at least in part to a successful outcome. Whilst we do not espouse such an approach in general, we nonetheless feel that it has it merits, and we feel that for this reason it is more natural to present our algorithm in line with the way in which it was conceived. We also hope that its correctness can be shown to be self-evident without recourse to consistency models and the like.

Finally a note on the naming of our algorithm and its utility. We chose `Concur' because as well as being a fragment of `concurrency', it is also an antonym of `differ'. This seemed appropriate given that algorithms such as ours are in some sense the opposite of those such as the `diff3' algorithm, itself recently formalised~\cite{diff3}. This algorithm will flag conflicts when attempting to merge changes. On the other hand our algorithm will always merge changes without conflicts, with the result that the resultant document may appear be nonsensical in places. This trade-off means that our algorithm and ones like it are hardly suitable for version control systems, however they find a use in real-time collaborative text editors, where the nonsensical parts can immediately be edited by users.

%% file: operationaltransformations.tex
\section{Operational transformations}
\label{section_operationaltransformations}

In this section we define our stringwise operational transformations. We do this informally first, taking two of the less obvious cases as examples, and then define them formally for each case. The main result of the section is that these definitions lead to the combined effect of any two operations executed sequentially being the same regardless of the order in which the operations are executed, provided that the second is suitably transformed relative to the first. Our operational transformations also preserve the intention of each individual operation, however we leave a proof of this until subsection~\ref{subsection_intentionpreservation}.

\begin{figure}[t]
\centering
\includegraphics[scale=0.5]{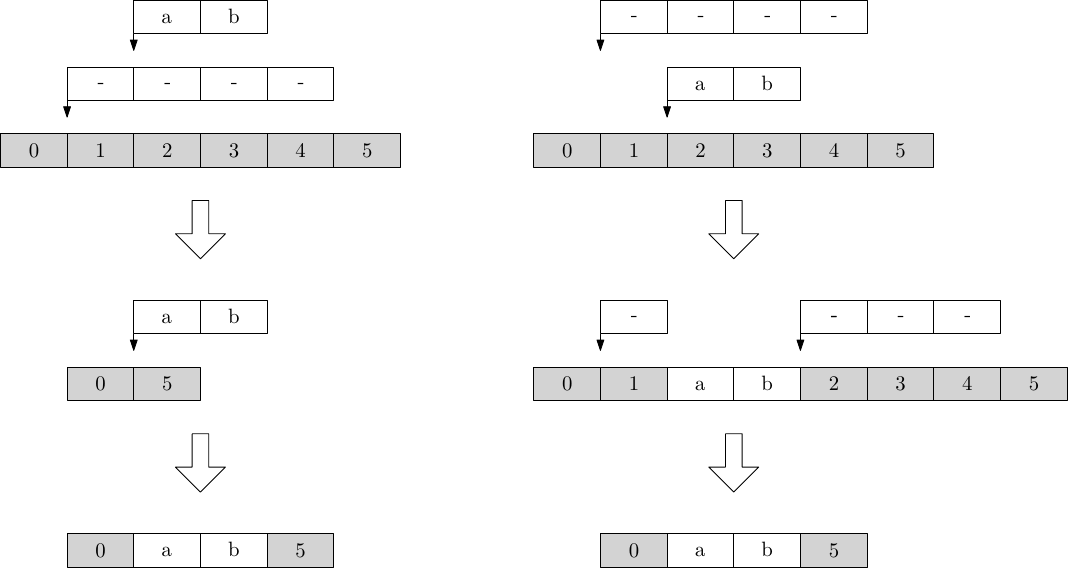}
\caption{Operational transformations when an insert splits a delete.}
\label{ot_insert_vs_delete}
\end{figure}

Consider then two users making concurrent changes to a document. The first deletes four characters, the second inserts two. After applying their own operations to their document, each user applies the other's. These operations need to transformed before being applied a second time if their effect is to be preserved, however. Figure~\ref{ot_insert_vs_delete} illustrates the requisite transformations. On the left, the insert must be moved one character to the left. On the right, the delete must be split in two, something we consider to be unavoidable if its intention is to be preserved. The splitting of deletes in cases like this sometimes leads to an increase in the number of operations as they are transformed, a process we call fragmentation. 

\begin{figure}[t]
\centering
\includegraphics[scale=0.5]{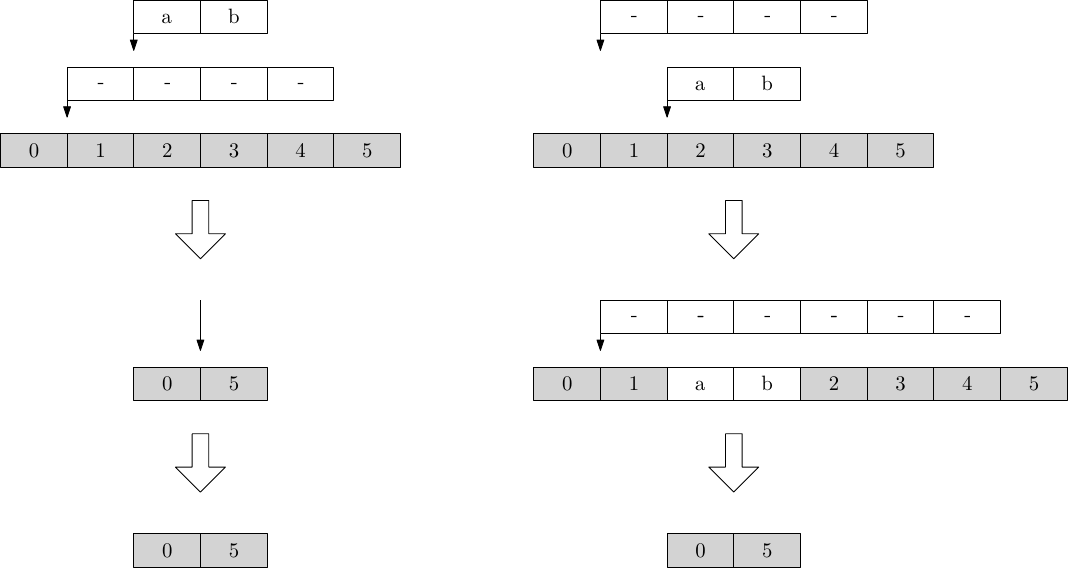}
\caption{Less than ideal operational transformations for an insert versus a delete.}
\label{ot_insert_vs_delete_compromised}
\end{figure}

One way to avoid fragmentation, adopted in~\cite{Ressel:1996:ITA:240080.240305} and in the majority of attempts since, is to consider each stringwise operation as a sequence of characterwise operations which cannot be split any further. We think this is wholly impractical. Another approach, adopted in~\cite{Cormack:1995:CCU:224964.225007}, is not to preserve the effect of the insert at all. Figure~\ref{ot_insert_vs_delete_compromised} illustrates these transformations. On the left, the insert is transformed into the empty operation. On the right, the transformed delete simply deletes the inserted string. This approach also mitigates against fragmentation because the transformed delete no longer has to be split in two, but at the expense of effectively throwing away the insert or, in other words, not preserving its intention. By contrast, our approach is not to compromise and to stick with what we call a comprehensive definition of stringwise operational transformations, even if this leads to fragmentation.

\begin{figure}[t]
\centering
\includegraphics[scale=0.5]{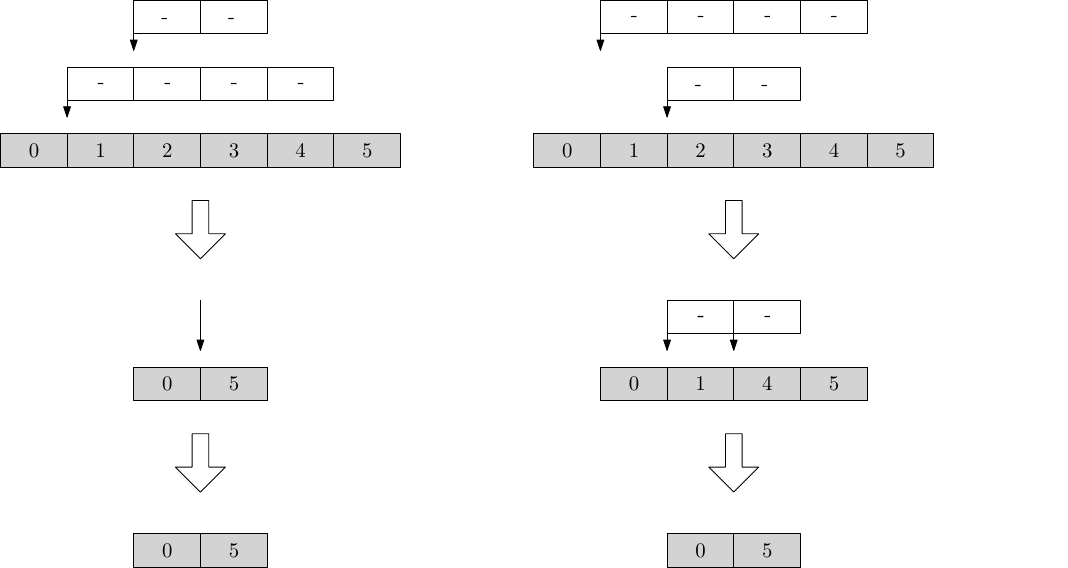}
\caption{Operational transformations when one delete splits another.}
\label{ot_delete_vs_delete}
\end{figure}

The other case that deserves mention is the case when one delete splits another. Figure~\ref{ot_delete_vs_delete} illustrates the transformations. On the left, the first delete when transformed becomes the empty operation, since all of the characters it was to delete have been deleted already. On the right, as in the first case, the transformed delete is split in two, however the two resulting deletes lie immediately next to each other once the other delete is applied, and can therefore be treated as one operation. This result has interesting consequences, in particular proving correctness in a more general setting turns out to be impossible without it.

Now let $\tau$ and $\rho$ be two arbitrary operations. We define $\tau\backslash\rho$ as the operation or operations that result when $\tau$ is transformed relative to $\rho$, and vice versa for $\rho\backslash\tau$. We can then state the combined effect of any two operations executed sequentially being the same regardless of the order in which they are executed, provided that the second is suitably transformed relative to the first, as the following equivalence:
\begin{equation}
\label{equivalence_effect}
\tau;\rho\backslash\tau\equiv\rho;\tau\backslash\rho
\end{equation}
In the remainder of this section we give formal definitions of operational transformations for all cases, define what is meant by two operations or sequences of operations being equivalent, then prove that this equivalence always holds.
\begin{definition}
Let $\Sigma$ be a non-empty, finite set of characters from some alphabet. A string is any finite sequence of characters from $\Sigma$, ranged over by $s$, $s'$ and so on. The length of a string $s$, written $|s|$, is the length of this sequence. The set of these strings is written $\Sigma^{*}$ and the set of non-empty strings $\Sigma^{+}$. We define the substring $s[n...m]$ to be the string formed by taking the $n$'th to the $m-1$'th characters of the string $s$ inclusive. We also make use of the abbreviations $s[...m]=s[0...m]$ and $s[n...]=s[n...|s|]$. We write $s'+s''$ for the concatenation of strings $s'$ and $s''$ in the usual sense of the word.
\end{definition}
We define the syntax of operations as follows:
\begin{definition}
The operations $\tau$, $\rho$ and so on range over the following set:
\[
\{i(n,s)|n\in\mathbb{N},s\in\Sigma^{+}\}\cup\{d(n,l)|n\in\mathbb{N},l\in\mathbb{N}^{+}\}
\]
\end{definition}
\begin{definition}
The operation $\epsilon$ ranges over the following set:
\[
\{e()\}
\]
\end{definition}
Intuitively $i(n,s)$ is an insert, $d(n,l)$ a delete and $e()$ is the operation that does nothing, otherwise known as the empty operation. We say that inserts and deletes have position $n$.

We define the effects of operations as follows:
\begin{definition}
$i(n,s)$, $d(n,l)$ and $e()$ are partial functions, defined only for suitable strings, in which case we have:
\[
\begin{myarray}[1pt]{rl}
\begin{myarray}[1pt]{rcl}
i(n,s'):\\
\vphantom{\vspace{2em}}
\end{myarray}
&
\begin{myarray}[1pt]{rcl}
\Sigma^{*}&\longrightarrow&\Sigma^{+}\\
s&\longmapsto&s[...n]+s'+s[n...]
\end{myarray}
\vspace{1em}
\\
\begin{myarray}[1pt]{rcl}
d(n,l):\\
\vphantom{\vspace{2em}}
\end{myarray}
&
\begin{myarray}[1pt]{rcl}
\Sigma^{+}&\longrightarrow&\Sigma^{*}\\
s&\longmapsto&s[...n]+s[n+l...]
\end{myarray}
\vspace{1em}
\\
\begin{myarray}[1pt]{rcl}
e():\\
\vphantom{\vspace{2em}}
\end{myarray}
&
\begin{myarray}[1pt]{rcl}
\Sigma^{*}&\longrightarrow&\Sigma^{*}\\
s&\longmapsto&s
\end{myarray}
\end{myarray}
\]
By a suitable string $s$ we mean $n\leqslant|s|$ in the case of inserts, $n+l\leqslant|s|$ in the case of deletes and any string in the case of the empty operation.
\end{definition}
\begin{definition}
Two single operations are equivalent, that is $\tau\equiv\rho$, if and only if $\tau(s)=\rho(s)$ for any string $s$ suitable for both $\tau$ and $\rho$.
\end{definition}
So equivalence is defined in terms of the effect of the operations in question. It is easy to check that $\tau\equiv\rho$ precisely when $\tau$ and $\rho$ are identical syntactically.

\begin{figure}[t]
\centering
\includegraphics[scale=0.75]{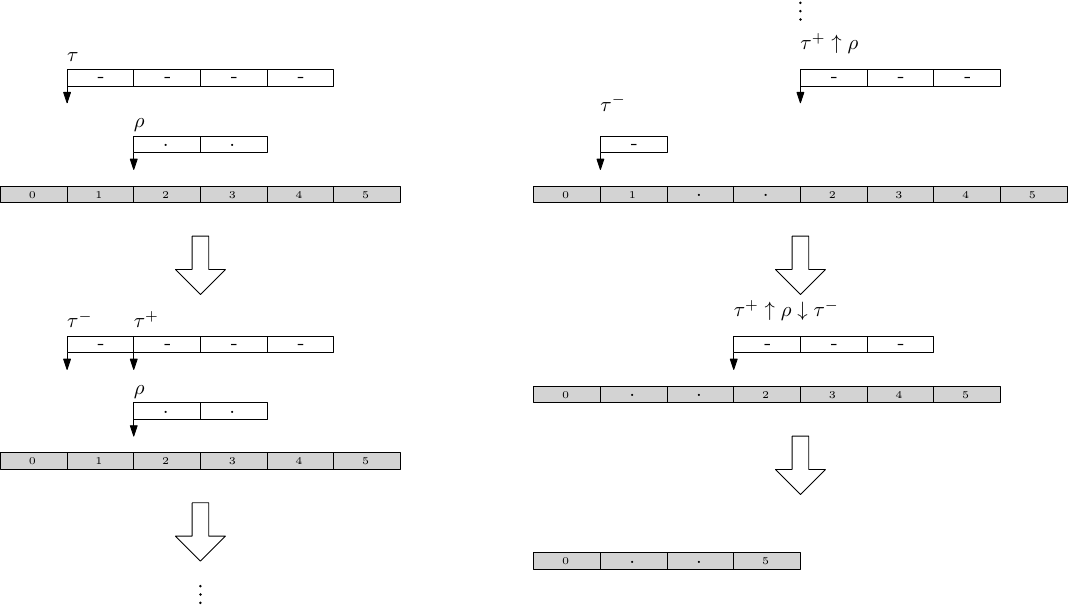}
\caption{The reasoning behind the equivalence $\tau\backslash\rho\equiv\tau^-;\tau^+\uparrow\rho\downarrow\tau^-$ when $\tau$ covers $\rho$.}
\label{ot_delete_covers_insert}
\end{figure}

We next extend the notion of equivalence to sequences of operations.
\begin{definition}
\label{definition_equivalence_sequences}
Consider two sequences of operations $\tau_1;\tau_2...;\tau_m$ and $\rho_1;\rho_2...;\rho_m$. We define them as being equivalent, that is $\tau_1;\tau_2...;\tau_n\equiv\rho_1;\rho_2...;\rho_n$, if an only if $\tau_m(...\tau_2(\tau_1(s)))=\rho_n(...\rho_2(\rho_1(s)))$ for any suitable string $s$. By suitable we mean not only that $\tau_1(s)$ and $\rho_1(s)$ are defined, but also $\tau_2((\tau_1(s))$, $\rho_2(\rho_1(s))$ and so on.
\end{definition}
Before continuing we make two points. The first point is that the notions of effect and equivalence here have nothing to do with the meaning of the underlying content. We hope it goes without saying that any treatment concerned with preserving meaning of this content, however this meaning might be defined, is a treatment of an entirely different problem to the one solved here. The second point is really an excuse for the definitions and results that follow. They are laborious, however a faithful implementation requires them. We nonetheless encourage the disinterested reader to move on to the next section.
\begin{definition}
\label{definition_corners}
\[
\begin{myarray}[10pt]{rl}
\lcorner{i(n,s)}=n&i(n,s)\hspace{-1.5pt}\raisebox{-1.5pt}{\rcorner}=n+|s|-1\\
\lcorner{d(n,l)}=n&d(n,l)\hspace{-1.5pt}\raisebox{-1.5pt}{\rcorner}=n+l-1
\end{myarray}
\]
\end{definition}
We call $\lcorner\tau$ and $\tau\xrcorner$ the corners of $\tau$, taking the right corner to be the position of the last character underneath the operation, so to speak. For example, in figure~\ref{ot_delete_covers_insert} the left and right corners of the delete are 1 and 4, respectively.
\begin{definition}
\label{definition_relative_positions}
\[
\begin{myarray}[3pt]{rcl}
\tau\lsep\rho&\text{iff}&\tau\xrcorner<\xlcorner\rho\\
\tau<\rho&\text{iff}&(\lcorner\tau<\xlcorner\rho)\wedge(\tau\xrcorner\geqslant\xlcorner\rho)\\
\tau\simeq\rho&\text{iff}&(\lcorner\tau=\xlcorner\rho)\wedge(\tau\neq\rho)\\
\tau>\rho&\text{iff}&(\lcorner\tau\leqslant\rho\rcorner)\wedge(\tau\xrcorner>\rho\rcorner)\\
\tau\rsep\rho&\text{iff}&\lcorner\tau>\rho\rcorner
\end{myarray}
\]
\end{definition}

These definitions formalise the idea of one non-empty operation overlapping another, regardless of whether the operations are inserts or deletes. Intuitively $\tau\simeq\rho$ when $\tau$ and $\rho$ start in the same place but are not equal, $\tau<\rho$ when $\tau$ starts to the left of $\rho$ but they overlap, and $\tau\lsep\rho$ when $\tau$ starts to the left of $\rho$ but they do not overlap. Similarly for $\tau>\rho$ and $\tau\rsep\rho$.
\begin{lemma}
For any two differing operations $\tau$ and $\rho$, exactly one of the relations in definition~\ref{definition_relative_positions} will hold.\qed
\end{lemma}
Next we define a series of partial transformations for inserts and deletes that formalise the idea of one non-empty operation being shifted one way or the other by another.
\begin{definition}
\label{definition_shift_insert}
For two inserts $i(n_1,s_1)$ and $i(n_2,s_2)$ with $n_1\geqslant n_2$:
\[
i(n_1,s_1)\uparrow i(n_2,s_2)=i(n_1+|s_2|,s_1). 
\]
For a delete $d(n_1,l_1)$ and insert $i(n_2,s_2)$ with $n_1\geqslant n_2$:
\[
d(n_1,l_1)\uparrow i(n_2,s_2)=d(n_1+|s_2|,l_1). 
\]
\end{definition}
\begin{definition}
\label{definition_shift_delete}
For two deletes $d(n_1,l_1)$ and $d(n_2,l_2)$ with $n_1\geqslant n_2+l_2$:
\[
d(n_1,l_1)\downarrow d(n_2,l_2)=d(n_1-l_2,l_1)
\]
For an insert $i(n_1,s_1)$ and delete $d(n_2,l_2)$ with $n_1\geqslant n_2+l_2$:
\[
i(n_1,s_1)\downarrow d(n_2,l_2)=i(n_1-l_2,s_1)
\]
\end{definition}
Intuitively $\tau\uparrow\rho$ is $\tau$ shifted to the right by the length of $\rho$ when $\rho$ is an insert; and $\tau\downarrow\rho$ is $\tau$ shifted to the left by the length of $\rho$ when $\rho$ is a delete. Note the restrictions on the relative positions of the operations in each case. There is never a need to shift an operation to the right by the length of an insert if that operation is already to its left. Similarly there is never a need to shift an operation to the left by the length of a delete unless that operation is to its right.

Finally we define partial transformations that split or crop one non-empty operation relative to another. The motivation for these can be seen in figure~\ref{ot_delete_covers_insert} again. Compare this with figure~\ref{ot_insert_vs_delete}, where the various steps involved in transforming the delete relative to the insert were left to the imagination. Figure~\ref{ot_delete_covers_insert} on the other hand makes these steps explicit. The delete is first subdivided, and then the right side must be further shifted twice. The result, as expected, is that the transformed delete operation will delete the same characters, albeit either side of the inserted characters, that the original delete operation would have deleted were the insert operation not to be applied first. Its intention is preserved, in other words.
\begin{definition}
\label{definition_delete_shift}
For two deletes $d(n_1,l_1)$ and $d(n_2,l_2)$:
\[
d(n_1,l_1)-d(n_2,l_2)=
\left\{
\begin{myarray}{lrlrl}
d(n_1,n_2-n_1)&n_1&<n_2&n_2<n_1&+\;l_1\leqslant n_2+l_2\\
d(n_2+l_2,n_1+l_1-n_2-l_2)&\;\;n_1+l_1&>n_2+l_2&n_2\leqslant&n_1<n_2+l_2
\end{myarray}
\right.
\]
\end{definition}
Intuitively if $\tau$ overlaps $\rho$ either to the left or the right, then $\tau-\rho$ is $\tau$ with that part overlapping with $\rho$ chopped off.
\begin{definition}
\label{definition_insert_chop}
For a delete $d(n_1,l_1)$ and an insert $i(n_2,s_2)$:
\[
\left.
\begin{myarray}{ll}
d(n_1,l_1)^{-}=d(n_1,n_2-n_1)\\
d(n_1,l_1)^{+}=d(n_2,l_1-n_2+n_1)
\end{myarray}
\right\}
\;n_1<n_2\;\;\;n_1+l_1>n_2+|s_2|
\]
\end{definition}
\begin{definition}
\label{definition_delete_chop}
For two deletes $d(n_1,l_1)$ and $d(n_2,l_2)$:
\[
\left.
\begin{myarray}{ll}
d(n_1,l_1)^{-}=d(n_1,n_2-n_1)\\
d(n_1,l_1)^{+}=d(n_2+l_2,n_1+l_1-n_2-l_2)
\end{myarray}
\right\}
\;n_1<n_2\;\;\;n_1+l_1>n_2+l_2
\]
\end{definition}
Intuitively if $\tau$ covers $\rho$, then $\rho$ splits $\tau$ into $\tau^{-}$ and $\tau^{+}$. If $\rho$ is an insert, the split takes place at the position of $\rho$ and none of $\tau$ is lost. If $\rho$ is a delete, only the parts of $\tau$ on either side of $\rho$ are kept. Note that we drop any reference to $\rho$ in these definitions, but it is always clear what $\rho$ is from the context.

We are now in a position to prove the main result of this section.
\begin{theorem}
\label{theorem_ot}
For any two single operations $\tau$ and $\rho$ and for suitable definitions of the transformed operations $\tau\backslash\rho$ and $\rho\backslash\tau$, equivalence~\ref{equivalence_effect} holds.
\begin{proof}
We break the proof down into cases.

When $\tau$ and $\rho$ are both inserts, we set $\tau=i(n_1,s_1)$, $\rho=i(n_2,s_2)$. Without loss of generality suppose that $n_1\leqslant n_2$. We treat the case when $n_1<n_2$ first, in which case we set $i(n_1,s_1)\backslash i(n_2,s_2)=i(n_1,s_1)$ and $i(n_2,s_2)\backslash i(n_1,s_1)=i(n_2,s_2)\uparrow i(n_1,s_1)$. For any suitable $s$ we then have:
\[
\begin{myarray}{rcl}
(i(n_1,s_1);i(n_2,s_2)\backslash i(n_1,s_1))(s)&=&(i(n_1,s_1);i(n_2+|s_1|,s_2))(s)\\
&=&i(n_2+|s_1|,s_2)(s[...n_1]+s_1+s[n_1...])\\
&=&s[...n_1]+s_1+s[n_1...n_2]+s_2+s[n_2...]\\
&=&i(n_1,s_1)(s[...n_2]+s_2+s[n_2...])\\
&=&(i(n_2,s_2);i(n_1,s_1))(s)\\
&=&(i(n_2,s_2);i(n_1,s_1)\backslash i(n_2,s_2))(s)
\end{myarray}
\]
For the sake of the reader who has gotten this far we omit similar proofs from now on. 

To continue, the case when $n_1=n_2$ is more subtle. We assume that there is some lexicographical ordering on the strings, that is for two strings $s_1,s_2\in\Sigma^{*}$ we have $s_1<s_2$, $s_1=s_2$ or $s_1>s_2$. When $s_1=s_2$ the transformations need do nothing for equivalence~\ref{equivalence_effect} to hold. Without loss of generality suppose now that $s_1<s_2$ and set $i(n_1,s_1)\backslash i(n_2,s_2)=i(n_1,s_1)$ and $i(n_2,s_2)\backslash i(n_1,s_1) = i(n_2+|s_1|,s_2)$. These transformations result in the lexicographically lesser of the two operations remaining in place whilst the other is shifted to the right, regardless of the order of application, and so again equivalence~\ref{equivalence_effect} holds. Hence in all the cases when both $\tau$ and $\rho$ are inserts, equivalence~\ref{equivalence_effect} holds.

From now on we just state the transformations in each case, leaving the proofs to interested readers. These are easily done along the lines of figure~\ref{ot_delete_covers_insert}.

When $\tau$ and $\rho$ are both deletes, we only need to consider the cases when $\tau=\rho$, $\tau\simeq\rho$ with $\lcorner\tau<\xlcorner\rho$, $\tau<\rho$ with $\tau\xrcorner<\rho\rcorner$, $\tau<\rho$ with $\tau\xrcorner=\rho\rcorner$, $\tau<\rho$ with $\tau\xrcorner>\rho\rcorner$, or $\tau\lsep\rho$. Symmetry takes care of the remaining cases. In the case when $\tau=\rho$, it suffices to set $\tau\backslash\rho=\rho\backslash\tau=\epsilon$. In the case when $\tau\lsep\rho$, it suffices to set $\tau\backslash\rho=\tau$ and $\rho\backslash\tau=\rho\downarrow\tau$. The other cases are a little more involved. In the case when $\tau\simeq\rho$ and $\tau\xrcorner<\rho\rcorner$ we set $\tau\backslash\rho=\epsilon$ and $\rho\backslash\tau=(\rho-\tau)\downarrow\tau$. In the case when $\tau<\rho$ with $\tau\xrcorner<\rho\rcorner$ we set $\tau\backslash\rho=\tau-\rho$ and $\rho\backslash\tau=(\rho-\tau)\downarrow\tau$ again. In the case when $\tau<\rho$ with $\tau\xrcorner=\rho\rcorner$ we set $\tau\backslash\rho=\tau-\rho$ and $\rho\backslash\tau=\epsilon$. Only the case when $\tau$ covers $\rho$ remains, namely when $\tau<\rho$ with $\tau\xrcorner>\rho\rcorner$. Here we set $\tau\backslash\rho=\tau^{-};\tau^{+}\downarrow\rho\downarrow\tau^{-}$
 and $\rho\backslash\tau=\epsilon$, noting again that the first of these transformations results in a single transformed operation, not two. Hence in all the cases when both $\tau$ and $\rho$ are deletes, equivalence~\ref{equivalence_effect} holds.

When $\tau$ is a delete and $\rho$ an insert, we need to consider the cases $\tau\rsep\rho$, $\tau>\rho$, $\tau\simeq\rho$, $\tau<\rho$, or $\tau\lsep\rho$. The relative positions of the right corners of the operations do not come into account. In the cases when $\tau\rsep\rho$, $\tau>\rho$, or $\tau\simeq\rho$ we set $\tau\backslash\rho=\tau\uparrow\rho$ and $\rho\backslash\tau=\rho$. In the case when $\tau<\rho$ we set $\tau\backslash\rho=\tau^{-};\tau^{+}\uparrow\rho\downarrow\tau^{-}$, as illustrated in figure~\ref{ot_delete_covers_insert}, and $\rho\backslash\tau=\rho\downarrow\tau^{-}$. Finally, in the case when $\tau\lsep\rho$ we set $\tau\backslash\rho=\tau$ and $\rho\backslash\tau=\rho\downarrow\tau$. Hence in all the cases when $\tau$ is a delete and $\rho$ an insert, equivalence~\ref{equivalence_effect} holds.

The trivial observation that if $\tau$ is an insert and $\rho$ a delete we simply swap the symbols above completes the proof.
\end{proof}
\end{theorem}

%% file: protocol.tex
\section{The protocol}
\label{section_protocol}

In this section we devise an algorithm that is able to keep any number of copies of a document in line whilst changes are made to them concurrently. In order to do so it utilises the method outlined in the previous section to transform one sequence of operations relative to another, together with a simple protocol which sits on top of the HTTP protocol. Since the HTTP protocol is based on the client-server model, so is our algorithm. Crucially, the choice of a client-server model leads to the correct application of operational transformations to both newly generated \emph{and} pending operations. We explain what we mean by this later. 

Just as in the previous sections, the treatment here is predominantly theoretical although we draw parallels with the workings of the implementation where appropriate. One instance where practical considerations had a bearing on the theory was the problem of latency. Initially this was overlooked, and then when issues arose, they had to be addressed. In the end the solution required no more than a refinement of the protocol. 

We therefore break this section into two. In the first part, we adopt a woolly notion of global time and present a protocol based on that, neglecting the problem of latency. This allows us to get the salient points across. In the second part, we do away with global time and adopt Lamport's ``happens before'' relation~\cite{lamport1978time}. This provides the correct context in which to explain the solution to the problem of latency, namely a refined protocol, and to give a general proof.

\begin{figure}[t]
\centering
\includegraphics[scale=0.75]{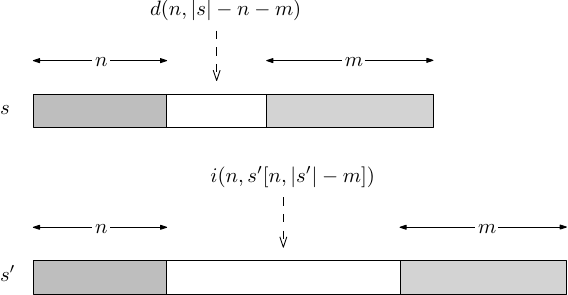}
\caption{Generating operations from a comparison of strings $s$ and $s'$.}
\label{generate_operations}
\end{figure}

Before getting going we briefly describe how stringwise operations are generated from a comparison of two differing strings. Such a comparison results in at most one delete and one insert operation. Figure~\ref{generate_operations} illustrates this. Here the shaded parts of strings $s$ and $s'$ are identical front and back. The middle parts, if there are any, contribute to a delete operation in the case of $s$ and an insert operation in the case of $s'$. It is easy to check the following:
\[
s'=(d(n,|s|-n-m);i(n,s'[n,|s'|-m]))(s)
\]
Note that it makes no difference whether the difference between the strings is one character or many, still at most two operations are generated. This is in marked contrast to characterwise operations, where copying a whole block of text into the input field, for example, might result in an overwhelming number of operations. The clear advantage of stringwise operations should be self-evident here.

To begin the first part of this section proper we note that the client-server model consists of any number of clients, we kick off with two for illustrative purposes, and a single server. Here each client has a copy of the document, as does the server, which also has a store of pending operations for each client. Communication is carried out by way of transactions, with each transaction consisting of two parts: a request from the client, which garners a response from the server. A request consists of a command and an optional sequence of operations. A response consists of either copies of the document or sequences of operations. The protocol consists of just three types of transaction, summarised as follows:

\begin{itemize}
\item$\mathsf{\scriptstyle INITIALISE}$: the server responds with a copy of its document,
\item$\mathsf{\scriptstyle PUT}$: the client puts a sequence of operations on the server, the server confirms,
\item$\mathsf{\scriptstyle GET}$: the client requests its pending operations, and the server duly responds.
\end{itemize}

\begin{figure}[t]
\centering
\includegraphics[scale=1]{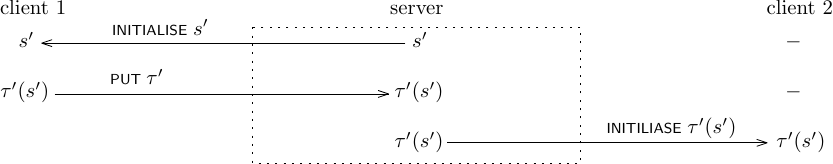}
\caption{Two $\mathsf{\scriptstyle INITIALISE}$ transactions with one $\mathsf{\scriptstyle PUT}$ transaction in-between}
\label{concur_initialise}
\end{figure}

Figure~\ref{concur_initialise} illustrates the likely first few transactions of a session. Here the first client is initialised with document $s'$ and then puts a sequence of operations $\tau'$ on the server, possibly over the course of several transactions. The server applies these operations to its own copy of the document and when a second client is initialised, it receives this amended document $\tau'(s')$. In this simple case both clients end up with copies of the document that are in line with the server's copy.

Before going any further we describe these illustrations in detail and more importantly the assumptions inherent in them. We assume that the server was initialised at some point, before any of the clients. We also assume that transactions are completed, by which we mean that requests always garner responses. This cannot be guaranteed, of course, however fault tolerance can be built in for the occasions when transactions fail. See the end of the related work and conclusions section for the details. Moving on, since information only ever flows in one direction we do not show requests and responses as separate arrows, instead showing each transaction as an arrow in the appropriate direction and labelled with the requisite information. One assumption that we do not have to make is that transactions are handled sequentially by both the clients and the server. This can in fact be guaranteed in the implementation, and it means that the arrows in these illustrations never meet and never cross. We therefore draw them horizontally, assuming that transactions happen instantaneously and time unfolds incrementally. This is our woolly notion of global time.

Figure~\ref{concur_base} illustrates the continuation of the session, with both clients having the document $\tau'(s')=s$. The first client now puts another sequence of operations $\tau$ on the server, again possibly over the course of several transactions, and this time the server stores these for the second client. Thus the first client's operations $\tau$ become the second client's pending operations and the server's document becomes $\tau(s)$. Then the second client puts its own sequence of operations $\rho$ on the server and here come the crucial steps:

\vspace{8pt}
\noindent\emph{the second client's operations do not become the first client's pending operations without first being transformed relative to second client's own pending operations} 
\vspace{8pt}

Thus the first client's pending operations are $\rho\backslash\tau$ and not just $\rho$.

\vspace{8pt}
\noindent\emph{the server does not apply the second client's operations to its document without first transforming them relative to the second client's pending operations}
\vspace{8pt}

So the server applies the operations $\rho\backslash\tau$ to its document, which becomes $(\tau;\rho\backslash\tau)(s)$. 

\vspace{8pt}
\noindent\emph{the second client's pending operations are also transformed relative to the operations it has just generated}
\vspace{8pt}

Therefore the second client's pending operations $\tau$ become $\tau\backslash\rho$. 

In this case both clients again end up with copies of the document that are in line with the server's copy, if we assume that $(\tau;\rho\backslash\tau)(s)=(\rho;\tau\backslash\rho)(s)$, an assumption based on theorem~\ref{theorem_equivalence_commutative_general}. Next we clear up a technicality used in these arguments, namely that it makes no difference whether a sequence of operations is put on the server by a particular client over the course of several transactions or just one:

\begin{figure}[t]
\centering
\includegraphics[scale=1]{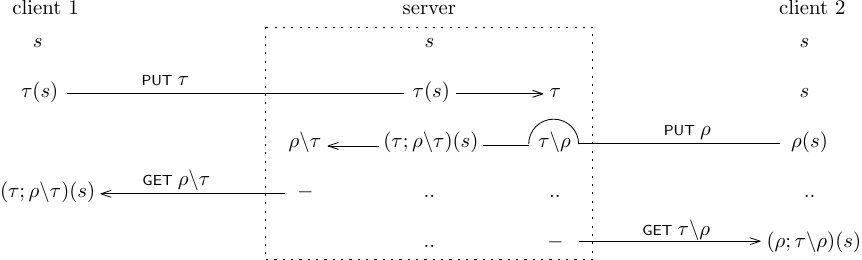}
\caption{Two $\mathsf{\scriptstyle PUT}$ transactions followed by two $\mathsf{\scriptstyle GET}$ transactions}
\label{concur_base}
\end{figure}

\begin{lemma}
Suppose the second client puts a sequence of operations $\rho$ on the server over the course of $n$ transactions, thus $\rho=\rho_1;\rho_2...;\rho_n$. Then its pending operations, if $\tau$ beforehand, become $\tau\backslash\rho$; the server's document, if $\tau(s)$ beforehand, becomes $(\tau;\rho\backslash\tau)(s)$; and the first client's pending operations, if the empty sequence beforehand, become $\rho\backslash\tau$.
\begin{proof}
By finite induction on the number of transactions. The base case is given by the steps above. Now suppose that the first $k-1$ transactions have taken place and set $\rho'=\rho_1;\rho_2...\rho_{k-1}$. By the induction hypothesis the second client's pending operations are $\tau\backslash\rho'$, the server's document $(\tau;\rho'\backslash\tau)(s)$ and the first client's pending operations $\rho'\backslash\tau$. Now suppose the client puts the next sequence of operations $\rho_k$ on the server. Then, for the second client's pending operations we have, by identity~\ref{identity_one_operation_transformed_relative_to_many}:
\[
(\tau\backslash\rho')\backslash\rho_k=\tau\backslash(\rho';\rho_k)
\]
For the server's document, by identity~\ref{identity_many_operations_transformed_relative_to_one}:
\[
(\tau;\rho'\backslash\tau;\rho_k\backslash(\tau\backslash\rho'))(s)=(\tau;(\rho';\rho_k)\backslash\tau)(s)
\]
And for the first client's pending operations, again by identity~\ref{identity_one_operation_transformed_relative_to_many}:
\[
\rho'\backslash\tau;\rho_k\backslash(\tau\backslash\rho')=(\rho';\rho_k)\backslash\tau
\]
This completes the proof.
\end{proof}
\end{lemma}
It should be clear that these arguments can be generalised but we leave off doing so until we have a proper notion of time. Nonetheless it is worth pointing out that the crux is here. Sequences of pending operations for each client must be held on the server because the server cannot pass them on to each client immediately. The HTTP protocol typically does not allow information to be pushed, only pulled, and so we simulated the pushing of information by having each client poll the server, a common practice. And the storage of pending operations on the server led in turn to them being transformed in the aforementioned symmetric way. 

We now come to the second part of this section and to an explanation of the solution to the problem of latency together with a general proof. Because there is no direct user interaction with the server and because it only handles one transaction at a time, the assumptions we have made about it thus far remain valid and, in particular, the storage and transformation of operations on the server does not need to change. On the other hand the state of a client may change due to user interaction whilst a transaction is in progress and this has not taken into account. We therefore do so now, bringing the treatment more into line with the implementation as it stands. This implementation includes a refined protocol consisting of just two types of transaction, summarised as follows:

\begin{itemize}
\item$\mathsf{\scriptstyle INITIALISE}$: the server responds with a copy of its document,
\item$\mathsf{\scriptstyle UPDATE}$: the client puts a sequence of operations on the server, and the server responds with the client's pending operations, suitably transformed.
\end{itemize}

Next we introduce the fact that clients keep not one copy of the document but two. The first is a working copy, the one formalised thus far, whilst the second is an editable copy considered to be the value of the input field made available to the user. Also from now on we work with an arbitrary, albeit fixed, number of clients, rather than just two. We represent the client involved in a particular transaction at any time as the $i$'th client, whilst any other client we represent as the $j$'th client.

\begin{figure}[t]
\centering
\includegraphics[scale=1]{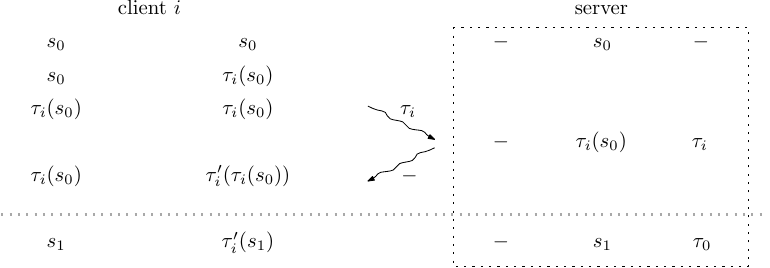}
\caption{The first $\mathsf{\scriptstyle UPDATE}$ transaction in the general case}
\label{general_first}
\end{figure}

Figure~\ref{general_first} illustrates the first $\mathsf{\scriptstyle UPDATE}$ transaction of a session with this new protocol, assuming that each client has already completed an $\mathsf{\scriptstyle INITIALISE}$ transaction. We describe these illustrations in detail again first. The far left column represents the $i$'th client's working copy of the document, the column next to this its editable copy. The server's columns are the same as before. We neglect to show the $j$'th client this time because it is not actively involved in the $i$'th client's transaction. Time unfolds top to bottom as before, but only does so during the course of a transaction. The topmost and bottommost lines show information across both the $i$'th client's and server's columns seemingly at the same time, but this only represents the fact that both have a state before and after the transaction. It is not a return to the woolly notion of global time.

Now suppose that a user makes a change to the $i$'th client's editable copy of the document. The $i$'th client duly updates its working copy and computes the requisite operations $\tau_i$, sending these to the server. Since this is the first $\mathsf{\scriptstyle UPDATE}$ transaction, the $j$'th client has no pending operations and so its pending operations become simply $\tau_i$. The server then updates its own copy of the document from $s_0$ to $\tau_i(s_0)$ and, since the $i$'th client also has no pending operations, returns nothing. So far this is all much the same as before. The difference now is that the user may have made further changes to the $i$'th client's editable copy of the document by the time the transaction is completed, therefore the editable copy becomes $\tau_i'(\tau_i(s_0))$. This means that there is no use in comparing the server's copy of the document with the $i$'th client's editable copy in any proof, rather the comparison has to be with its working copy. In this case it is easy to see that the two remain in line. Finally, we rename $\tau_i$ to $\tau_0$ and set $\tau_0(s_0)$ to $s_1$.

\begin{figure}[t]
\centering
\includegraphics[scale=1]{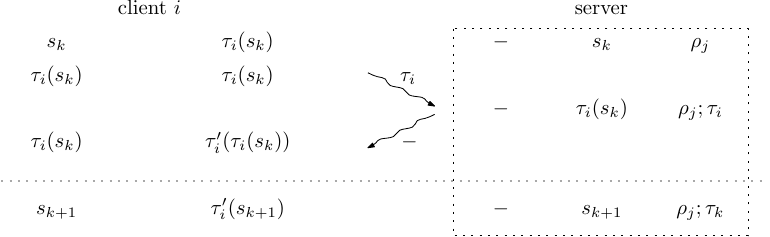}
\caption{The $k$'th $\mathsf{\scriptstyle UPDATE}$ transaction in the general case with no pending operations}
\label{general_kth_no_pending}
\end{figure}

We next consider the $k$'th $\mathsf{\scriptstyle UPDATE}$ transaction. There are two possibilities: either the $i$'th client completed the previous transaction and therefore has no pending operations, or it did not. Figure~\ref{general_kth_no_pending} illustrates the first possibility, with $\tau_i$ and $\tau_i'$ being re-used. Note that there is an important difference between this transaction and the first, namely that we cannot now assume that the $i$'th client's working and editable copies start off in line. In the case of the first $\mathsf{\scriptstyle UPDATE}$ transaction this could be assumed, because the implementation ensures that user interactions are discarded until the $\mathsf{\scriptstyle INITIALISE}$ transaction has completed, at which point the editable copy is set to be in line with the working copy. Now this is not the case and the difference $\tau_i$ between the editable and working copies could have come about either after completion of the previous transaction or whilst it was in progress. Either way it does not matter, however we draw attention to the fact that the topmost line showing the editable and working copies as being in line is missing, to make the point. Also note that should this $k$'th $\mathsf{\scriptstyle UPDATE}$ transaction be the second, we would equate the $\tau_i$ here with the previous $\tau_i'$. 

To continue, this transaction unfolds in a similar way to the first. Because the $j$'th client now has pending operations $\rho_j$, the $i$'th client's operations $\tau_i$ are appended to these. We then rename $\tau_i$ to $\tau_k$ and set $\tau_k(s_k)$ to $\tau_{k+1}$. Again the $i$'th client's working copy and the server's copy of the document, both being $s_{k+1}$, end up being in line if we assume that both start as being so. This assumption forms the induction hypotheis in an inductive proof based on the number of $\mathsf{\scriptstyle UPDATE}$ transactions, of which the first transaction is the base case. We present that proof once the theory is explained.

\begin{figure}[t]
\centering
\includegraphics[scale=1]{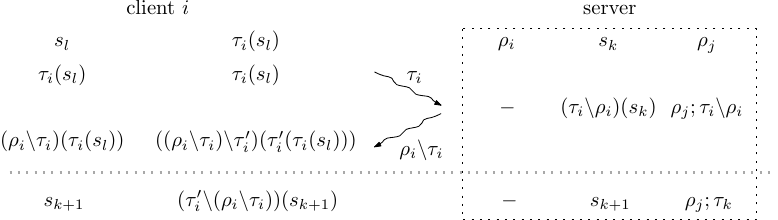}
\caption{The $k$'th $\mathsf{\scriptstyle UPDATE}$ transaction in the general case with pending operations}
\label{general_kth_pending}
\end{figure}

Now we come to the other possibility for the $k$'th $\mathsf{\scriptstyle UPDATE}$ transaction, namely that the $i$'th client did not initiate the previous transaction. Figure~\ref{general_kth_pending} illustrates this. Here the $i$'th client has pending operations $\rho_i$ and its working copy of the document $s_l$, for $l<k$, will be equal the server's copy of the document after the $l-1$'th $\mathsf{\scriptstyle UPDATE}$ transaction, namely the last transaction initiated by the $i$'th client itself. The following lemma will prove useful:
\begin{lemma}
\label{lemma_ith_working}
$\rho_i(s_l)=s_k$
\begin{proof}
We just have to observe that $\rho_i$ is $\tau_l;...\tau_{k-1}$ and since $\tau_l(s_l)=s_{l+1}$ all the way up to $\tau_{k-1}(s_{k-1})=s_k$, the result follows.
\end{proof}
\end{lemma}
As usual the operations $\tau_i$ the $i$'th client sends to the server are transformed relative to it's pending operations to become $\tau_i\backslash\rho_i$ before the server applies them to its copy of the document and appends them to the $j$'th client's pending operations $\rho_j$. And again, as usual rather than return the $i$'th client's pending operations directly, the server first transforms them relative to the operations $\tau_i$ to become $\rho_i\backslash\tau_i$. The $i$'th client duly applies them to its working copy of the document $s_l$ to give $(\rho_i\backslash\tau_i)(\tau_i(s_l))$. Now we rename not $\tau_i$ but $\tau_i\backslash\rho_i$ to $\tau_k$ and set $\tau_k(s_k)$ to $\tau_{k+1}$. In order to show that the $i$'th client's working copy and the server's copy of the document remain in line we have the following lemma:
\begin{lemma}
\label{lemma_general_kth_pending}
$(\rho_i\backslash\tau_i)(\tau_i(s_l))=s_{k+1}$
\begin{proof}
\[
\begin{myarray}[3pt]{rcl}
(\rho_i\backslash\tau_i)(\tau_i(s_l))&=&(\tau_i;\rho_i\backslash\tau_i)(s_l)\\
&=&(\rho_i;\tau_i\backslash\rho_i)(s_l)\\
&=&(\tau_i\backslash\rho_i)(\rho_i(s_l))\\
&=&(\tau_i\backslash\rho_i)(s_k)\\
&=&\tau_k(s_k)\\
&=&s_{k+1}\\
\end{myarray}
\]
Here we have made use of lemma~\ref{lemma_ith_working} and the usual identities.
\end{proof}
\end{lemma}
Lastly the $i$'th client must also apply its transformed pending operations $\rho_i\backslash\tau_i$ to its editable copy of the document. Further changes to this copy during the course of the transaction however means that it has become $\tau_i'(\tau_i(s_i))$ and therefore the pending operations must be transformed again, this time relative to $\tau_i'$, becoming $(\rho_i\backslash\tau_i)\backslash\tau_i'$, before being applied. The editable copy therefore becomes $((\rho_i\backslash\tau_i)\backslash\tau_i')(\tau_i'(\tau_i(s_l)))$, as illustrated in figure~\ref{general_kth_pending}. A proof along similar lines to that given in lemma~\ref{lemma_general_kth_pending} will equate this handful to $(\tau_i'\backslash(\rho_i\backslash\tau_i))(s_{k+1})$ but for the sake of the reader we omit this.

We end this section by summarising these results formally:
\begin{theorem}
\label{theorem_quiescence_fixed}
Consider a fixed number of clients, each having completed an $\mathsf{\scriptstyle INITIALISE}$ transaction. Then after any subsequent $\mathsf{\scriptstyle UPDATE}$ transaction, the working copy of the client that completed the transaction and the server's copy of the document are in line.
\begin{proof}
By induction on the number of $\mathsf{\scriptstyle UPDATE}$ transactions. Figure~\ref{general_first} illustrates the base case of the first $\mathsf{\scriptstyle UPDATE}$ transaction. Figures~\ref{general_kth_no_pending} and~\ref{general_kth_pending} illustrate that if, after the $k$'th $\mathsf{\scriptstyle UPDATE}$ transaction, the working copy of client that completed the transaction and the server's copy of the document are in line then, after $k+1$'th $\mathsf{\scriptstyle UPDATE}$ transaction, this is also true. So this holds after any number of $\mathsf{\scriptstyle UPDATE}$ transactions.
\end{proof}
\end{theorem}
Finally, we can drop the requirement that the number of clients be fixed:
\begin{theorem}
\label{theorem_quiescence}
Consider an increasing number of clients. Then after any transaction, the working copy of the client that completed the transaction and the server's copy of the document are in line.
\begin{proof}
Suppose a client completes an $\mathsf{\scriptstyle INITIALISE}$ transaction. Then its working, and indeed editable, copies are in line with the server's copy of the document when the $\mathsf{\scriptstyle INITIALISE}$ transaction completes. Moreover since the pending operations of the other clients and the server's copy of the document remain unchanged, after any subsequent $\mathsf{\scriptstyle UPDATE}$ transaction, the working copy of the client that completed the transaction and the server's copy of the document are in line by theorem~\ref{theorem_quiescence_fixed}. Should other clients complete an $\mathsf{\scriptstyle INITIALISE}$ transaction, the same argument can be used.
\end{proof}
\end{theorem}

Aside from the proof that our operational transformations preserve the intention of each individual operation, found in subsection~\ref{subsection_intentionpreservation}, the proofs in this section together with those in sections~\ref{section_operationaltransformations} and~\ref{section_decreasingdiagrams} give the first ever formally correct concurrency control algorithm for collaborative text editors.

%% file: consistency.tex
\section{Consistency}
\label{section_consistency}

In this section we prove that our algorithm is correct against the standard consistency model, although as we mentioned in the introduction, we hope that our algorithm's correctness has been shown to be self-evident without recourse to consistency models. Furthermore it is not unreasonable to remark that the standard consistency model can be a little problematic in places. We therefore modify it as we go along before drawing parallels.

Known as the CCI model, the standard consistency model requires the following three properties of a concurrency control algorithm to hold:
\begin{itemize}
\item \textbf{C}onvergence
\item \textbf{C}ausality, or precedence preservation
\item \textbf{I}ntention preservation
\end{itemize}
We tick these off one by one in what follows.

\subsection{Convergence}
\label{subsection_convergence}

To begin with we describe the concept of quiescence in the context of concurrency control algorithms. It is a time when there are no operations left to be executed by any client. A concurrency control algorithm is said to be convergent if it ensures that all the client's copies of the document are in line at quiescence, whenever this occurs. Immediately this is problematic, since there is no such thing as global time in a distributed system, as we know. We therefore redefine both quiescence and convergence, rather than relying on something akin to our own previous woolly notion of global time.

We first define what we call local quiescence. Consider the response part of any transaction. We know that the client no longer has pending operations on the server the moment the response is sent, and also that if it contains any pending operations they are immediately executed on the client the moment it is received. Therefore we define local quiescence as a combination of the moment on the server that the response is sent together with moment it happens before, namely the moment on the client when it is received. This allows us to define what we call local convergence as the property that the client's working copy and the server's copy of the document should be in line at local quiescence. So to say that our algorithm is locally convergent is no more than a restatement of theorem~\ref{theorem_quiescence}.

In choosing to compare the working copies of client's documents with the server's copy rather than their editable copies it may appear as if we are making a compromise. As mentioned in section~\ref{section_protocol}, however, it is impossible to do any better. Recall that the editable copy is considered to be no more than the value of the input field itself, therefore nothing can be said about it being in line with the server's copy at any given moment beyond remarking that if no user interactions take place during the course of a transaction, then the client's working and editable copies of the document will be in line when the transaction completes. Algorithms that execute operations the moment they are generated, as ours does, are called optimistic~\cite{nichols1995high}.

\begin{figure}[t]
\centering
\includegraphics[scale=0.75]{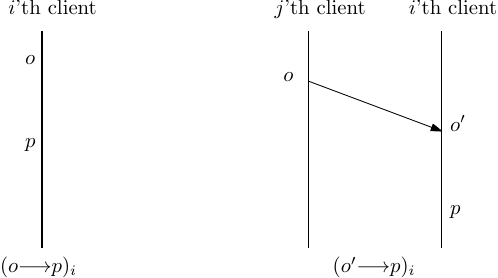}
\caption{The definition of $o\causes p$.}
\label{causality}
\end{figure}

Is there a more general proof that is closer in spirit to the standard consistency model's definition of convergence? The only moments in time that we have to work with, so to speak, are those moments on the server immediately a response is sent. If we take the pending operations the server has stored for every other client at any such moment and apply them to the working copies of these clients at the moments they completed their last transaction, it turns out not unsurprisingly that resulting copies are all in line. However we think that this is irrelevant, and do not give the proof. What we think is relevant is that over a series of transactions the working copy of the client that completes any transaction and the server's copy of the document are always in line at the moment the transaction is completed, and so we leave it at that.

\subsection{Causality or precedence preservation}
\label{subsection_causalitypreservation}
So far we have not mentioned causality, or precedence as it is also known, because it played no part in the development of our algorithm. Preserving causality is an issue for algorithms based on a peer-to-peer model, where the lack of a centralised controlling process makes keeping track of the order of operations extremely difficult as they are bandied about in all directions. By contrast we could say that our algorithm preserves causality a priori, since it has never been an issue. Nonetheless we give the standard definition of causality, which is based on Lamport's ``happens before'' relation, together with a proof that our algorithm preserves it.

To begin with we couch the generation and execution of operations on clients formally in terms of events. Operations are written $o$ and $p$, their transformed counterparts $o'$ and $p'$ respectively. We write $o_i$ to represent the event of operation $o$ being generated on the $i$'th client and $p'_j$, say, to represent the event of the transformed operation $p'$ being executed by the $j$'th client. We write $o_i\causes p_i$ if $o_i$ occurs before $p_i$, $o'_i\causes p_i$ if $o'_i$ occurs before $p_i$ and so on. Now we can give a definition of causality in terms of events:
\begin{definition}
The operation $o$ causes, or precedes, the operation $p$, written $o\causes p$, when either $o_i\causes p_i$ or $o'_i\causes p_i$.
\end{definition}
We abbreviate $o_i\causes p_i$ as $(o\causes p)_i$ and $o'_i\causes p_i$ as $(o'\causes p)_i$. Also $o'_i\causes p'_i$ can happen, which we abbreviate $(o'\causes p')_i$. So $o\causes p$ when either $(o\causes p)_i$ or $(o'\causes p)_i$. For an illustration that should make things clearer see figure~\ref{causality}. It should also be clear that if $o\causes p$ we cannot have $p\causes o$. This very valuable contribution, namely the realisation that Lamport's ``happens before'' relation on events leads to a relation on operations, is due to~\cite{Ellis:1989:CCG:66926.66963}. 

\begin{figure}[t]
\centering
\includegraphics[scale=0.75]{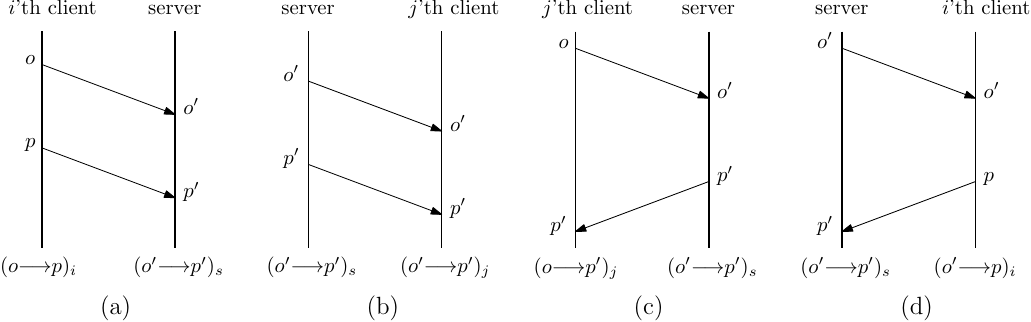}
\caption{Operations are always communicated in sequential order.}
\label{order_preservation}
\end{figure}

Now we come to the definition of the preservation of causality. An algorithm is said to preserve causality if, whenever $o\causes p$, $o$ is executed before $p$ on all clients. Formally:
\begin{definition}
An algorithm preserves causality when the following implications hold:
\[
\begin{myarray}{rcl}
(o\causes p)_i&\Rightarrow&(o'\causes p')_j\\
(o'\causes p)_i&\Rightarrow&(o'\causes p')_j\wedge\exists k\,(o\causes p')_k
\end{myarray}
\]
Here we assume as usual that $j\neq i$, and also that $i,j\neq k$.
\end{definition}
Now we formalise the notion of operations being put on the server. It is somewhat arbitrary whether we write $(o'\causes p')_s$ or $(o\causes p)_s$ here. We choose the former to emphasise the fact that operations are transformed the moment they arrive.
\begin{definition}
If $o$ is put on the server before $p$ we write $(o'\causes p')_s$.
\end{definition}
We next formalise the assumptions made in section~\ref{section_protocol} relating to the sequential order of operations on both client and server. Again for an illustration should make things clearer, see figure~\ref{order_preservation}.
\begin{assumption}
\label{assumption_clienttoserver}
If $(o\causes p)_i$ then $(o'\causes p')_s$.
\end{assumption}
\begin{assumption}
\label{assumption_servertoclient}
If $(o'\causes p')_s$ then $(o'\causes p')_j$.
\end{assumption}
\begin{assumption}
\label{assumption_serverandclient1}
If $(o\causes p')_i$ then $(o'\causes p')_s$ and vice versa.
\end{assumption}
\begin{assumption}
\label{assumption_serverandclient2}
If $(o'\causes p')_s$ then $(o'\causes p)_j$ and vice versa.
\end{assumption}
The proof that our algorithm preserves causality is now straightforward.
\begin{lemma}
\label{lemma_preserves_causality}
Our algorithm preserves causality.
\begin{proof}
We make use of figure~\ref{order_preservation}. If $(o\causes p)_i$ we join parts (a) and (b) to get $(o'\causes p')_j$. If $(o'\causes p)_i$ then by part (d) we must have $(o'\causes p')_s$. Then by part (b) we have $(o'\causes p')_j$ except for one client, since $o$ must be generated somewhere. Then by part (c) we have $(o\causes p')_j$ for one value of $j$, say $k$, that is $\exists k\,(o\causes p')_k$ and we are done.
\end{proof}
\end{lemma}

\subsection{Intention preservation} 
\label{subsection_intentionpreservation}
This property relates not to the workings of concurrency control algorithms as a whole but only to operational transformations. It turns out that precisely what it means for the intention of an operation to be preserved under transformation is not quite straightforward to formalise. However we jump through the hoops and prove that our stringwise operational transformations preserve intention in a way that hopefully appeals to common sense.

\begin{figure}[t]
\centering
\includegraphics[scale=0.5]{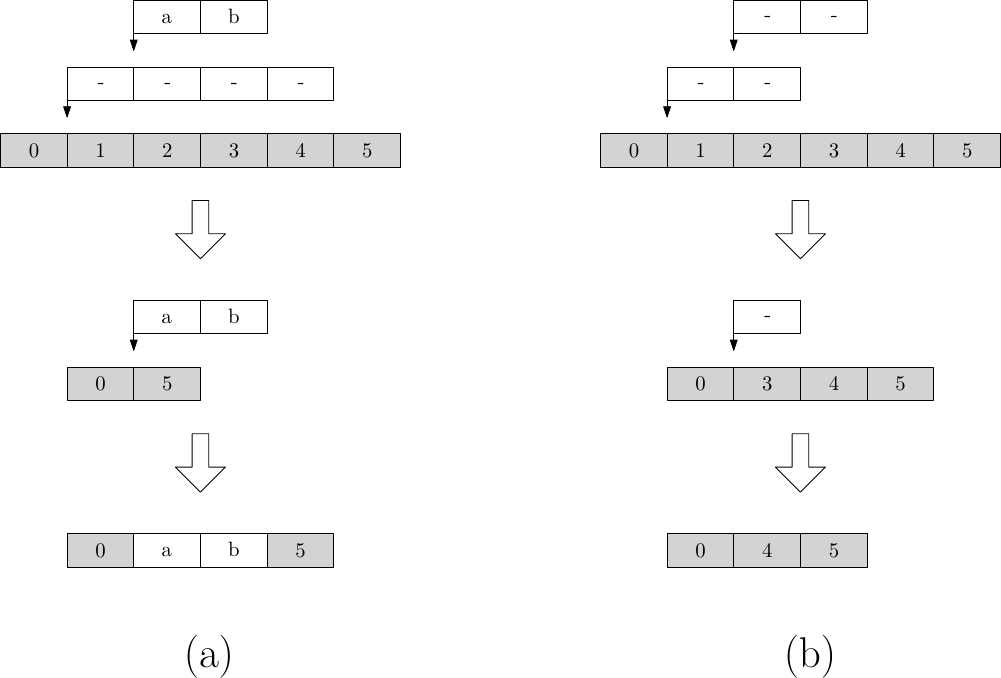}
\caption{Subtleties when transforming operations relative to deletes.}
\label{transformation_subtelties_1}
\end{figure}

Roughly speaking, the intention of an insert is preserved if, once transformed, it still inserts the same characters in the same place. There is one caveat, illustrated in figure~\ref{transformation_subtelties_1}~(a), namely that the characters cannot necessarily be inserted in the same place if the left hand corner of the insert overlaps the transforming delete. In this case the transformed insert does the best that it can, so to speak, inserting its characters immediately to the right of the deleted characters. Nonetheless it is reasonable to state that the intention of the insert is preserved. If we consider the string after both operations have been executed, we find that the inserted characters are where we would expect them to be. 

In a similar vein, roughly speaking the intention of a delete is preserved if, once transformed, it still deletes the same characters. Again there is a caveat, illustrated in figure~\ref{transformation_subtelties_1}~(b), namely that the characters cannot all be deleted if the transforming delete has already deleted some or all of them. This is easily accounted for by simply insisting that the transformed operation deletes only those characters that remain. Nonetheless again it is reasonable to state that the intention of the first delete is preserved, even though the second delete did part of the job for it, so to speak. If we consider the string after both operations have been executed, we find that the only requisite characters have been deleted.

We claim that our stringwise operational transformations always preserve the intention of operations. We have just given two of the subtler cases and, admittedly, there are others. Recall from section~\ref{section_operationaltransformations}, for example, the transformation of a delete by an insert which results in the delete being split in two. It is easy to check that the intention of the delete is preserved in this case, however. It does, after all, delete the same characters after the transformation as it would have done before. This is illustrated in figure~\ref{ot_insert_vs_delete}.

We could leave it at that, however a proof is needed. So we begin with the formalising the intent of operations and hope that it clarifies rather than occludes:
\begin{definition}
The intention of an insert is an ordered pair consisting of the index of the character immediately to the left of which the insert operation's characters are to be inserted or zero if there are no characters, together with a string of the insert operation's characters themselves. The intention of a delete is a set of the indexes of the characters it deletes:
\[
\begin{myarray}{rcl}
\llbrack i(n,s)\rrbrack&=&(n,s)\\
\llbrack d(n,l)\rrbrack&=&\{n,...n+l-1\}
\end{myarray}
\]
\end{definition}
Now we employ a little sleight of hand in order to ease the formalism that follows. Looking at figure~\ref{transformation_subtelties_1}~(a), we see that the character with index 5 keeps this index after the execution of the first operation, and again, after the second operation. The characters in figure~\ref{transformation_subtelties_1}~(b) also keep their original indexes in this way. In fact this has always been the case in these illustrations, see figures~\ref{ot_insert_vs_delete}, \ref{ot_insert_vs_delete_compromised} and~\ref{ot_delete_vs_delete} in section~\ref{section_operationaltransformations}, for example.

When the transforming operation is an insert this renumbering makes the preservation of intention easy to formalise. For the transformation of an insert $i(n,s)$ relative to another insert $i(n',s')$, it should be clear that if we allow characters to keep their original indexes we have:
\[
\llbrack i(n,s)\backslash i(n',s')\rrbrack=(n,s)
\]
Similarly for the transformation of a delete $(n,l)$ relative to an insert $i(n',s')$, again it should be clear that if we allow characters to keep their original indexes we have:
\[
\llbrack d(n,l)\backslash i(n',s')\rrbrack=\{n,...n+l-1\}
\]
And so in all cases when $\tau$ is an arbitrary operation and $\rho$ is an insert we have:
\begin{equation}
\llbrack\tau\backslash\rho\rrbrack=\llbrack\tau\rrbrack
\label{equation_intention_preservation_operations_vs_insert}
\end{equation}
When the transforming operation is a delete we know that the cases can be more subtle. For the transformation of an insert $i(n,s)$ relative to a delete $d(n',l')$ we have:
\[
\llbrack i(n,s)\backslash d(n',l')\rrbrack=
\left\{
\begin{myarray}[4pt]{rl}
(n'+l',s)&n'\leqslant n<n'+l'\\
(n,s)&\textrm{otherwise}
\end{myarray}
\hspace{0.25em}
\right.
\]
In other words, if the left hand corner of the insert overlaps the delete, the insert is effectively moved immediately to the delete operation's right. See the illustration on left hand side of figure~\ref{ot_insert_vs_delete} again. With a little care we can re-use the formalism of section~\ref{section_operationaltransformations}:
\begin{equation}
\llbrack\tau\backslash\rho\rrbrack=
\left\{
\begin{myarray}[4pt]{rl}
\llbrack\tau\uparrow\rho^{+}\rrbrack&\tau\simeq\rho\vee\tau>\rho\\
\llbrack\tau\rrbrack&\textrm{otherwise}
\end{myarray}
\hspace{0.25em}
\right.
\label{equation_intention_preservation_insert_vs_delete}
\end{equation}
For the transformation of a delete $d(n,l)$ relative to another delete $d(n',l')$ we can do better. If the deletes do not overlap, the transformed delete will delete the same characters, otherwise it will only delete the characters left by the transforming delete. Formally:
\[
\llbrack d(n,l)\backslash d(n',l')\rrbrack=\{n,..n+l-1\}\backslash\{n',..n'+l'-1\}
\]
So in the case of both $\tau$ and $\rho$ being deletes we have:
\begin{equation}
\llbrack\tau\backslash\rho\rrbrack=\llbrack\tau\rrbrack\backslash\llbrack\rho\rrbrack
\label{equation_intention_preservation_delete_vs_delete}
\end{equation}
And so we have a proof of sorts.
\begin{lemma}
Our stringwise operational transformations preserve intention.
\begin{proof}
It is straightforward to check that in all cases that equalities~\ref{equation_intention_preservation_operations_vs_insert}, \ref{equation_intention_preservation_insert_vs_delete} and~\ref{equation_intention_preservation_delete_vs_delete} hold.
\end{proof}
\end{lemma}

%% file: conclusions.tex
\section{Related work and conclusions}

To the best of our knowledge ours is the first formally correct concurrency control algorithm for collaborative text editors and it is reasonable to ask why, given the considerable amount of literature surrounding this problem. See~\cite{otfaq} for a partial synopsis, for example, and~\cite{journals/cscw/LiL10} for the following negative view: ``proofs are very complicated and error-prone...we can only conclude that an algorithm achieves convergence but cannot draw any conclusion about intention preservation.'' 

\begin{figure}[t]
\centering
\includegraphics[scale=0.875]{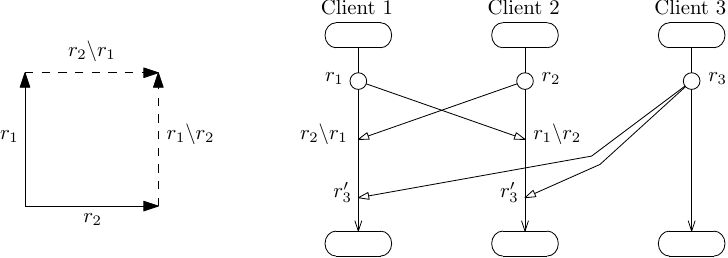}
\caption{Transformation properties TP1 and TP2}
\label{tp}
\end{figure}

Our work contradicts this view. For example we appear to be the first in proving that our operational transformations preserve intention, with more than one recent work~\cite{li2005commutativity,weiss:inria-00275754} agreeing with our own findings that this has never been done before. In doing so, however, we run the risk of occluding with formalism what is intuitively pretty obvious. When proving convergence on the other hand the formalism feels a lot less strained, in fact we think that the theorems in sections~\ref{section_operationaltransformations},~\ref{section_decreasingdiagrams} and~\ref{section_protocol} are genuinely useful in demonstrating that our algorithm works perfectly in practice. We also hope that the crux of the argument, illustrated in figure~\ref{concur_base}, is perfectly clear. The rest, as they say, then follows.

It is perhaps a little surprising then that stringwise operational transformations such as ours have never been adopted. In our opinion the reason is that they lead to a convergence problem requiring a novel proof as opposed to a simple inductive one. As mentioned earlier in section~\ref{section_operationaltransformations}, less than ideal stringwise operational transformations have been used in the past in order to admit inductive convergence proofs, but as a consequence they do not preserve intention~\cite{Cormack:1995:CCU:224964.225007}. Stringwise operational transformations have occasionally cropped up elsewhere~\cite{Sun96aconsistency} but the details are vague. Another work~\cite{Sun98achievingconvergence} does indeed acknowledge the effects of what we call fragmentation, but there are no proofs and apparently the implementation led to ``complications''. Lastly a more recent work~\cite{shao2009optimized} uses stringwise operations but splitting deletes is avoided, apparently in order to ``simplify presentation and stay focused on the main contribution''.

On the other hand, in spite of their limitations in our view, characterwise operational transformations have always been used. Despite the fact that they mitigate against fragmentation, in our opinion making any consistency proof considerably easier, nonetheless a convincing proof against the standard consistency model for an algorithm that uses them seems never to have emerged. Any such proof must start with the result that the operational transformations themselves are correct, and such proofs are lacking from the earliest attempts~\cite{Ellis:1989:CCG:66926.66963,Ressel:1996:ITA:240080.240305,Sun98achievingconvergence,suleiman1998concurrent}, indeed the operational transformations used in these attempts have all been found to be incorrect in~\cite{imine2003proving}. 

This work also contains what appears to be the first set of characterwise operational transformations that have been proved correct using an automatic theorem prover~\cite{spike}. By way of comparison we attenuate our own stringwise operational transformations to characterwise ones:
\[
\begin{myarray}{cc}
i(n,c)\backslash i(m,d)=
\left\{
\begin{myarray}{ll}
i(n,c)&\quad n<m\\
i(n,c)&\quad n=m\wedge c<d\\
i(n,c)&\quad n=m\wedge c=d\\
i(n+1,c)&\quad n=m\wedge c>d\\
i(n+1,c)&\quad n>m\\
\end{myarray}
\right.
&\quad
i(n,\_)\backslash d(m)=
\left\{
\begin{myarray}{ll}
i(n,\_)&\quad n\leqslant m\\
i(n-1,\_)&\quad n>m
\end{myarray}
\right.
\end{myarray}
\]
\[
\begin{myarray}{cc}
d(n)\backslash i(m,\_)=
\left\{
\begin{myarray}{ll}
d(n)&\quad n<m\\
d(n+1)&\quad n\geqslant m
\end{myarray}
\right.
&\quad
d(n)\backslash d(m)=
\left\{
\begin{myarray}{ll}
d(n)&\quad n<m\\
e()&\quad n=m\\
d(n-1)&\quad n>m
\end{myarray}
\right.
\end{myarray}
\]
Aside from two small differences these agree with those given in~\cite{imine2003proving}. The first of these is that their algorithm tries to differentiate between inserts with the same position by keeping track of their original position before resorting to a lexicographical ordering on the characters. This is perhaps a little over-engineered. The second is that when both inserts are identical, their algorithm transforms one of the two inserts into the empty operation. We think this is a move towards transformations being in some way bound to the meaning of the underlying content, and cannot agree with it. However these are small gripes.

\begin{figure}[t]
\centering
\includegraphics[scale=0.75]{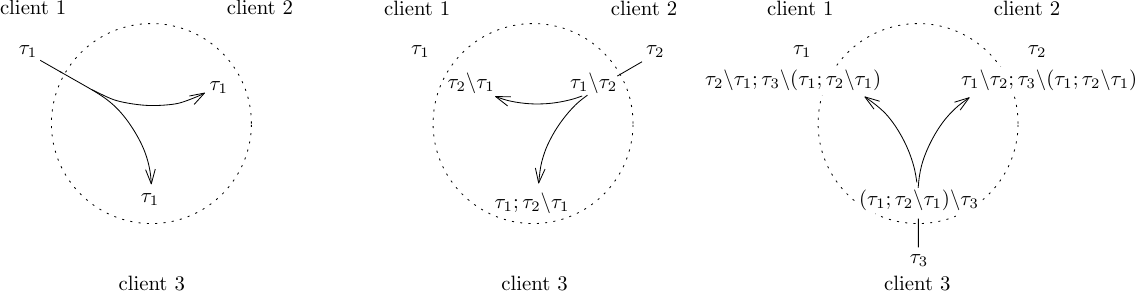}
\caption{Imposing a total ordering on concurrent operations}
\label{tp2_client_server}
\end{figure}

Moving on, with the problem of finding a correct set of characterwise operational transformations apparently solved, why then did a convergence proof still remain elusive? In our opinion one of the main reasons is a predilection for the peer-to-peer model that has continued from the early days~\cite{Ellis:1989:CCG:66926.66963,Cormack:1995:CCU:224964.225007} into recent times~\cite{DBLP:journals/tpds/LiL07,10.1109/TPDS.2009.173} whereas implementations based on the client-server model have always been rare~\cite{nichols1995high}. The peer-to-peer model brings with it the disadvantage of the lack of a centralised controlling process and, as a consequence, the problem of imposing a total ordering on the operations for the purposes of transformation~\cite{Vidot:2000:CCD:358916.358988,DBLP:journals/tpds/LiL07}, or making do without one. With operations executed concurrently any imposed total ordering is bound to be somewhat arbitrary, however without one the problem becomes harder. In these cases the operational transformations have to satisfy not only our own equivalence~\ref{equivalence_effect}, which has always been known as transformation property TP1~\cite{Ressel:1996:ITA:240080.240305}, but also the mysterious transformation property TP2~\cite{Ressel:1996:ITA:240080.240305}. We reproduce the commonplace illustrations of these properties in figure~\ref{tp}, including the first because of its resemblance to our own decreasing diagrams. 

We give an explanation of this mysterious TP2. Whilst TP1 relates to the effect of operations on the underlying document, TP2 requires that operational transformations must ensure that the combined effect of two operations, not on the underlying document but on a third concurrent operation, must be the same regardless of their order, given of course that the second is transformed relative to the first:
\[
\tau_3\backslash(\tau_1;\tau_2\backslash\tau_1)=\tau_3\backslash(\tau_2;\tau_1\backslash\tau_2)
\]
Incidentally, whether our own operational transformations satisfy this property is moot. Because our algorithm is based on a client-server model rather than a peer-to-peer one, it is able to impose a total ordering on operations for the purposes of transformation by virtue of the fact that transactions are handled sequentially. We show a simple case of how this total ordering is imposed in figure~\ref{tp2_client_server}, where $\tau_1<\tau_2<\tau_3$ on the server. Here pending operations are shown inside the dotted circles whilst operations already executed on the clients are shown outside. Note that the ordering $\tau_1<\tau_2$ does not mean that $\tau_2$ is never executed before $\tau_1$ on some clients, transformed or otherwise. At client 2, for example, $\tau_2$ will be executed before $\tau_1\backslash\tau_2$. What it does mean, however, is that on the server $\tau_2$ never occurs before $\tau_1$, transformed or otherwise, and since transformations always happen on the server, when any other operation is transformed relative to these two operations, it is always transformed relative to $\tau_1$ before $\tau_2$, again transformed or otherwise. Thus we see $\tau_3\backslash(\tau_1;\tau_2\backslash\tau_1)$, but would never see $\tau_3\backslash(\tau_2;\tau_1\backslash\tau_2)$. 

\begin{figure}[t]
\centering
\includegraphics[scale=1]{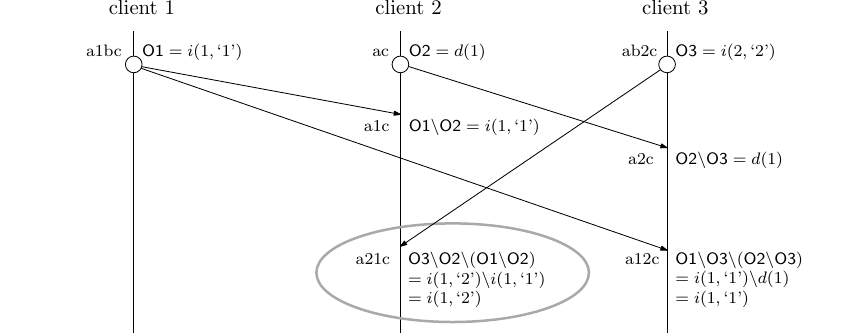}
\caption{The amended false-tie puzzle}
\label{false_tie}
\end{figure}

\subsubsection*{The false tie puzzle}

We called the transformation property TP2 mysterious and we give the reasons why. We look at what is called the false tie puzzle~\cite{Sun:1997:GOT:266838.267366}, which could be said to test the correctness of operational transformations when a total ordering on operations for the purposes of transformation has not been imposed. We reproduce the commonplace illustration of this puzzle in figure~\ref{false_tie}, missing out some of the operations executed at the first client because they are not in fact needed for the complete puzzle.

The ``puzzle'' is that the algorithm diverges because of what is known as a false tie, a seemingly incorrect operational transformation that results from two inserts occupying the same position:
\[
i(1,\text{`2'})/i(1,\text{`1'})=i(1,\text{`2'})
\]
Note that this operational transformation is different from our own, as it leaves the lexicographically greater of the two inserts in place rather than the lexicographically lesser when the two inserts are tied. In fact our operational transformation would not break the puzzle but this is not the point. The operational transformation above is just as valid, with the choice of whether to leave the lexicographically lesser or greater insert in place when two inserts are tied being an arbitrary one. The point is that this puzzle cannot differentiate between correct and incorrect sets of operational transformations. What is wrong is not the operational transformations themselves but the algorithm implicit in the puzzle itself.

It is perhaps not surprising that one solution to this puzzle is to require that the operational transformations satisfy TP2. Comparing the sequences of operations executed by the second and third clients, rewriting on occasion and employing TP1 where necessary we get:
\[
\begin{myarray}{rcl}
O2;O1\backslash O2;O3\backslash O2\backslash(O1\backslash O2)&\equiv&O3;O2\backslash O3;O1\backslash O3\backslash(O2\backslash O3)\\
...&\equiv&O2;O3\backslash O2;O1\backslash O3\backslash(O2\backslash O3)\\
O1\backslash O2;O3\backslash O2\backslash(O1\backslash O2)&\equiv&O3\backslash O2;O1\backslash O3\backslash(O2\backslash O3)\\
O1\backslash O2;(O3\backslash O2)\backslash(O1\backslash O2)&\equiv&...\\
O3\backslash O2;(O1\backslash O2)\backslash(O3\backslash O2)&\equiv&O3\backslash O2;O1\backslash O3\backslash(O2\backslash O3)\\
(O1\backslash O2)\backslash(O3\backslash O2)&\equiv&O1\backslash O3\backslash(O2\backslash O3)\\
O1\backslash O2\backslash(O3\backslash O2)&\equiv&...\\
O1\backslash(O2;O3\backslash O2)&\equiv&O1\backslash(O3;O2\backslash O3)
\end{myarray}
\]
This begs the question, are there any operational transformations that satisfy TP2? Certainly none of those studied in~\cite{ImineTCS06}, only the characterwise ones outlined in~\cite{imine2003proving}, the ones closely in agreement with our own attenuated stringwise operational transformations. We also agree with~\cite{ImineTCS06} that it is a difficult if not impossible task to verify whether a set of operational transformations satisfies TP2 without the help of an automated theorem prover. There are simply too many cases to consider. We wonder at this point whether our own stringwise operational transformations do so but then, as we have pointed out, the matter is moot.